\documentclass[twoside] {siamltex}
\usepackage{graphicx}
\usepackage{cite}
\usepackage{amsmath}
\title{Linnik point spread functions, time-reversed logarithmic diffusion equations, and blind deconvolution of electron microscope imagery}{
\author {Alfred S. Carasso and Andras E. Vladar \thanks
{Applied and Computational Mathematics Division, National Institute of Standards and Technology, Gaithersburg, MD 20899.
(alfred.carasso@nist.gov),} and\\
{Microsystems and Nanotechnology Division,
National Institute of Standards and Technology, Gaithersburg, MD 20899. (andras.vladar@nist.gov).}}
\begin{document}
\maketitle
\begin{abstract}
	A non-iterative direct blind deconvolution procedure, previously used successfully to sharpen Hubble Space Telescope imagery, is now found useful
        in sharpening nanoscale scanning electron microscope (SEM) and helium ion microscope (HIM) images. The method is restricted to images $g(x,y)$,
        whose Fourier transforms $\hat{g}(\xi,\eta)$ are such that $log~|\hat{g}(\xi,0)|$ is globally monotone decreasing and convex. The method
        is not applicable to defocus blurs. A point spread function in the form of a Linnik probability density
        function is postulated, with parameters obtained by least squares fitting the Fourier transform of the {\em preconditioned} microscopy image.
        Deconvolution is implemented in {\em slow motion} by marching backward in time, in Fourier space, from $t=1$ to $t=0$, in an associated logarithmic
        diffusion equation.
        Best results are usually found in a {\em partial deconvolution} at time $\bar{t}$, with $0 < \bar{t} < 1$, rather than in total deconvolution at $t=0$.
        The method requires familarity with microscopy images, as well as interactive search for optimal parameters.



\end{abstract}

\pagestyle{myheadings}
\thispagestyle{plain}

\section{Introduction}
There are numerous processes for the extraction of meaningful information contained in various
images. For proper identification, interpretion, and analysis, a human observer must first be able to perceive and recognize all of the information contained in a  given image.
Often, valuable pertinent information lies in faint details. A combination of local contrast enhancement and sharpening of fine
image details, is especially helpful for scanning electron microscope (SEM) and helium ion microscope (HIM) images.
Modern SEMs and HIMs can resolve sub-nanometer details \cite{Vlad}, but their images often suffer from low contrast and appreciable noise.
As a result, fine details can easily be overlooked. Denoising, contrast stretching, and sharpening, can significantly improve the results of quantitative analyses of these images.

This paper  describes a new {\em non-iterative}, direct blind deconvolution procedure for sharpening images obtained from scanning electron microscopes, and Helium ion microscopes. As was the case in \cite{{carAPEX},{carSEM}, {carOP}, {carIM}}, the present method is only applicable to a restricted class of blurred images $g(x,y)$, with Fourier transforms $\hat{g}(\xi,\eta)$ such that $log~|\hat{g}(\xi,0)|$ is globally monotone decreasing and convex. The method
does not apply to defocus and motion blurs.
We view the given image $g(x,y)$ as the convolution of the desired sharp image $f(x,y)$ with an unknown point spread function $h(x,y)$,
together with an unknown amount of noise $n(x,y)$,
\begin {equation}
g(x,y)=\int_{R^2} h(x-u, y-v)f(u,v)dudv +n(x,y) =h\otimes f +n(x,y).
\label{eq:0}
\end{equation}
After an appropriate {\em preconditioning} of the microscopy image $g(x,y)$, the method is based on postulating a point spread function in the form of a Linnik probability density function $h(x,y)$, and then identifying the 
parameters $\gamma, ~\lambda > 0,$~ in the corresponding Linnik optical transfer function 
\begin{equation}
\hat{h}(\xi,\eta)=\{1+4 \pi^2 \gamma (\xi^2 + \eta^2)\}^{-\lambda},
	\label{eq:1.01}
\end{equation}
by least squares fitting the Fourier transform of the 
preconditioned $g(x,y)$.  Here, for any function $p(x,y)$ in $L^1(R^2)$, we define its Fourier transform $\hat{p}(\xi,\eta)$ by
\begin{equation}
\hat{p}(\xi,\eta)= \int_{R^2} p(x,y) e^{-2\pi i(\xi x +\eta y)}dx dy.
\label{eq:1.02}
\end{equation}
With the Linnik optical transfer function $\hat{h}(\xi,\eta)$ in Eq.~(\ref{eq:1.01}), it is useful to define $\hat{h}(\xi,\eta,t)$ as follows for $0 \leq t \leq 1$,
\begin{equation}
\hat{h}(\xi,\eta,t)=\{1+4 \pi^2 \gamma (\xi^2 + \eta^2)\}^{-\lambda t}, \qquad 0 \leq t \leq 1.
        \label{eq:1.03}
\end{equation}
\begin{figure}
        \centerline{\includegraphics[width=4.5in]{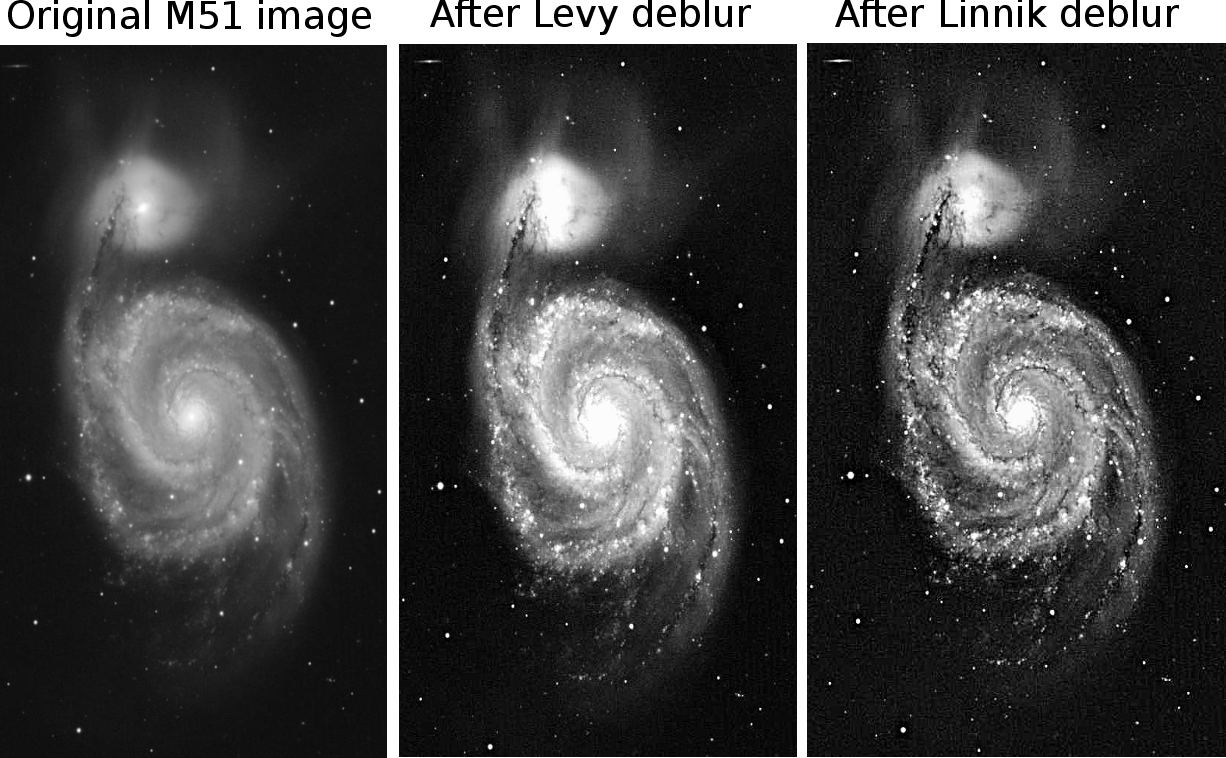}}
        \caption{As shown in \cite{carIM}, the use of Linnik probability densities, rather than L\'{e}vy stable densities, produces better results in deblurring
astronomical imagery. Above Whirlpool Galaxy image was obtained at Kitt Peak National Observarory. Successful Linnik deblurring of several Hubble Space Telescope color images may also be found in \cite{carIM}.
        Horizontal Field Width (HFW) in above images is approximately 53,000 light years.}
\end{figure}

Previous non-iterative direct blind deconvolution methods, based on candidate point spread functions in the form of  heavy-tailed L\'{e}vy stable probability densities, were successfully used in several applications
\cite{carSEB, carDIR, carAPEX, carSEM,carOP}. In that approach, fast Fourier transform (FFT) algorithms are used to implement the deconvolution as a backward in time stepwise marching procedure, from $t=1$ to $t=0$,
in a  diffusion equation involving fractional powers of the negative Laplacan. Stopping the process prior to reaching $t=0$,  produces a {\em partial deconvolution} which is often beneficial.
In \cite{carIM}, the use of Linnik probability densities, rather than L\'{e}vy stable densities, was found to produce significantly better results in deblurring
Hubble Space Telescope and other astronomical imagery. This is illustrated in Figure 1.1, in the case of a Kitt Peak National Observatory image of the Whirlpool Galaxy (M51), obtained by Rector and Ramirez.
In the Linnik blind deconvolution procedure discussed in \cite{carIM}, deconvolution unfolds as a backward in time marching procedure, from $t=1$ to
$t=0$, in a diffusion equation involving the {\em logarithm} of the identity plus
the negative Laplacian, $ w_t=-\lambda\{log(I+\gamma(-\Delta)\} w$. 
Using the Lipschitz exponent theory developed in \cite{carLIP}, it can be shown {\em quantitatively} that the rightmost image in Figure 1.1 is significantly sharper than the other two images. The behavior of Linnik versus L\'{e}vy optical transfer functions at high and low frequencies, is discussed in \cite[Sections 4-6]{carIM}, and that behavior is used to explain these improved sharpness results.

\section{Preconditioning microscopy images}
An important first step prior to Linnik blind deconvolution of microscopy images, consists of applying adaptive
histogram equalization (ADHE) to the image. However, while this brings out useful information, it also produces
significant noise. Smoothing that noisy image by convolution with a low exponent L\'{e}vy probability density
function is helpful. This is illustrated in Figure 2.1. In the smoothed ADHE image $g(x,y)$, least squares fitting $log |\hat{g}(\xi,0)|$ with the expression 
$-\lambda(log(1+4 \pi^2 \gamma \xi^2) -b$, where $b > 0$ is an appropriately chosen constant, leads to parameter values for $\lambda$ and $\gamma$
        in  $\hat{h}(\xi,\eta)$ defined in Eq.~(\ref{eq:1.01}). This is shown in Figure 2.2.
\begin{figure}
        \centerline{\includegraphics[width=4.5in]{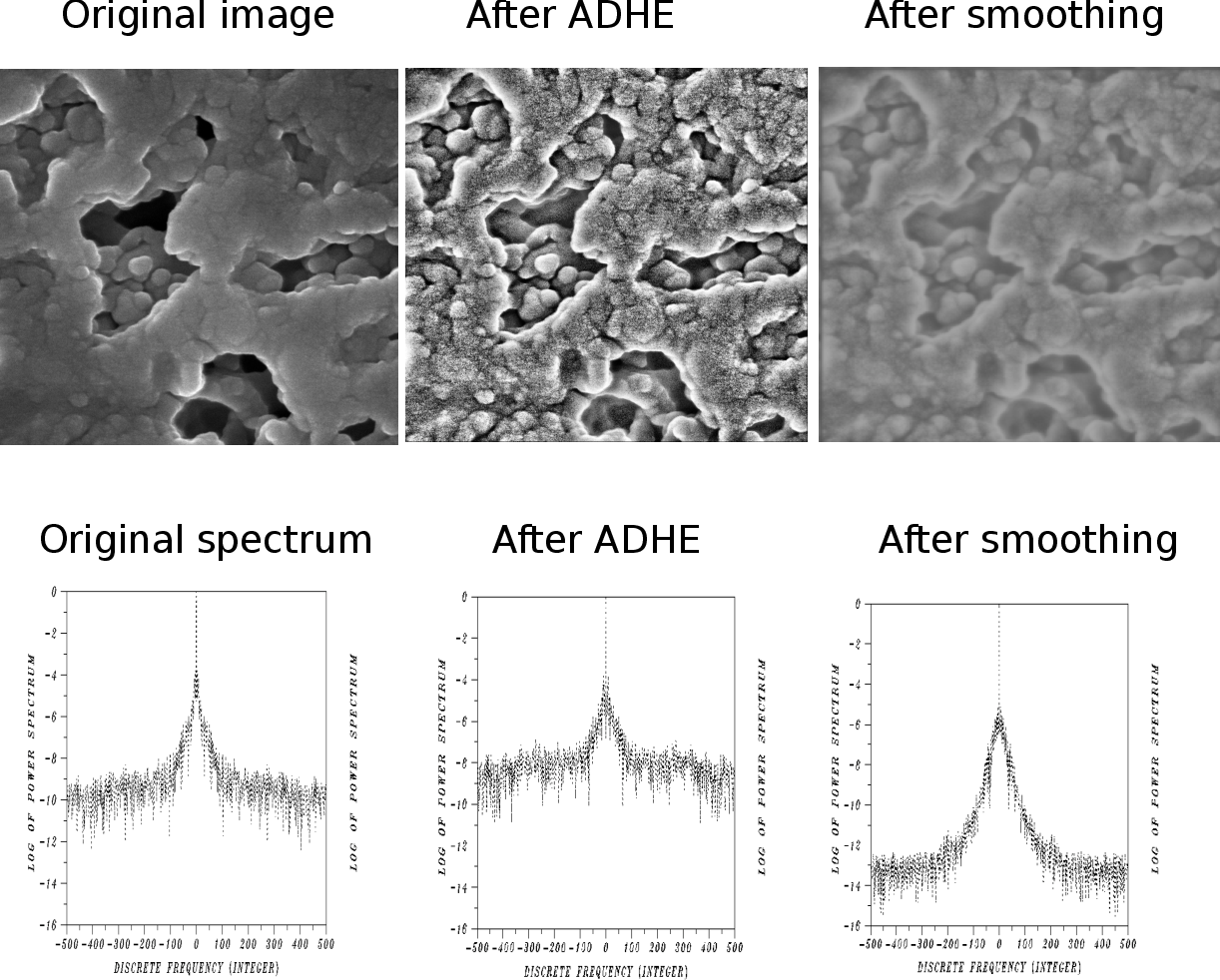}}
	\caption{Adaptive Histogram Equalization (ADHE), reveals valuable information while generating significant noise that must be smoothed out.
	This is reflected in the respective Fourier spectra. Above images are $1 \mu m$ HFW secondary electron images.}
\end{figure}
\begin{figure}
        \centerline{\includegraphics[width=4.4in]{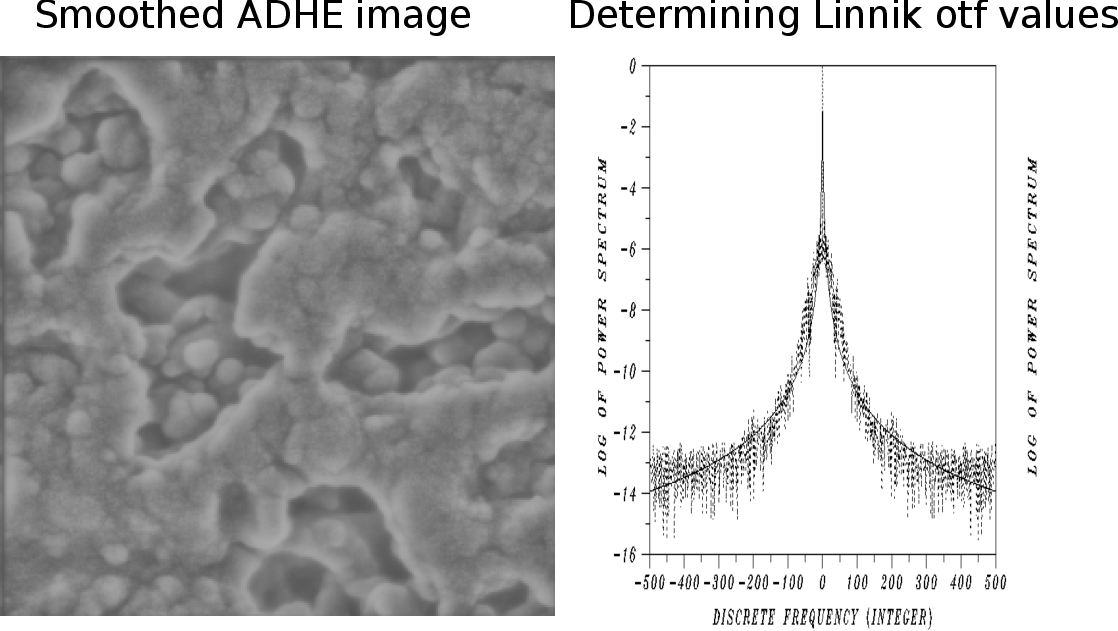}}
	\caption{In the smoothed ADHE image $g(x,y)$, least squares fitting $log |\hat{g}(\xi,0)|$ with the expression $-\lambda(log(1+4 \pi^2 \gamma \xi^2) -1.5$, leads to parameter values $\lambda=0.969,~\gamma=1.64$,
	for $\hat{h}(\xi,\eta)$ in Eq.~(\ref{eq:1.01})}
\end{figure}
\begin{figure}
        \centerline{\includegraphics[width=4.25in]{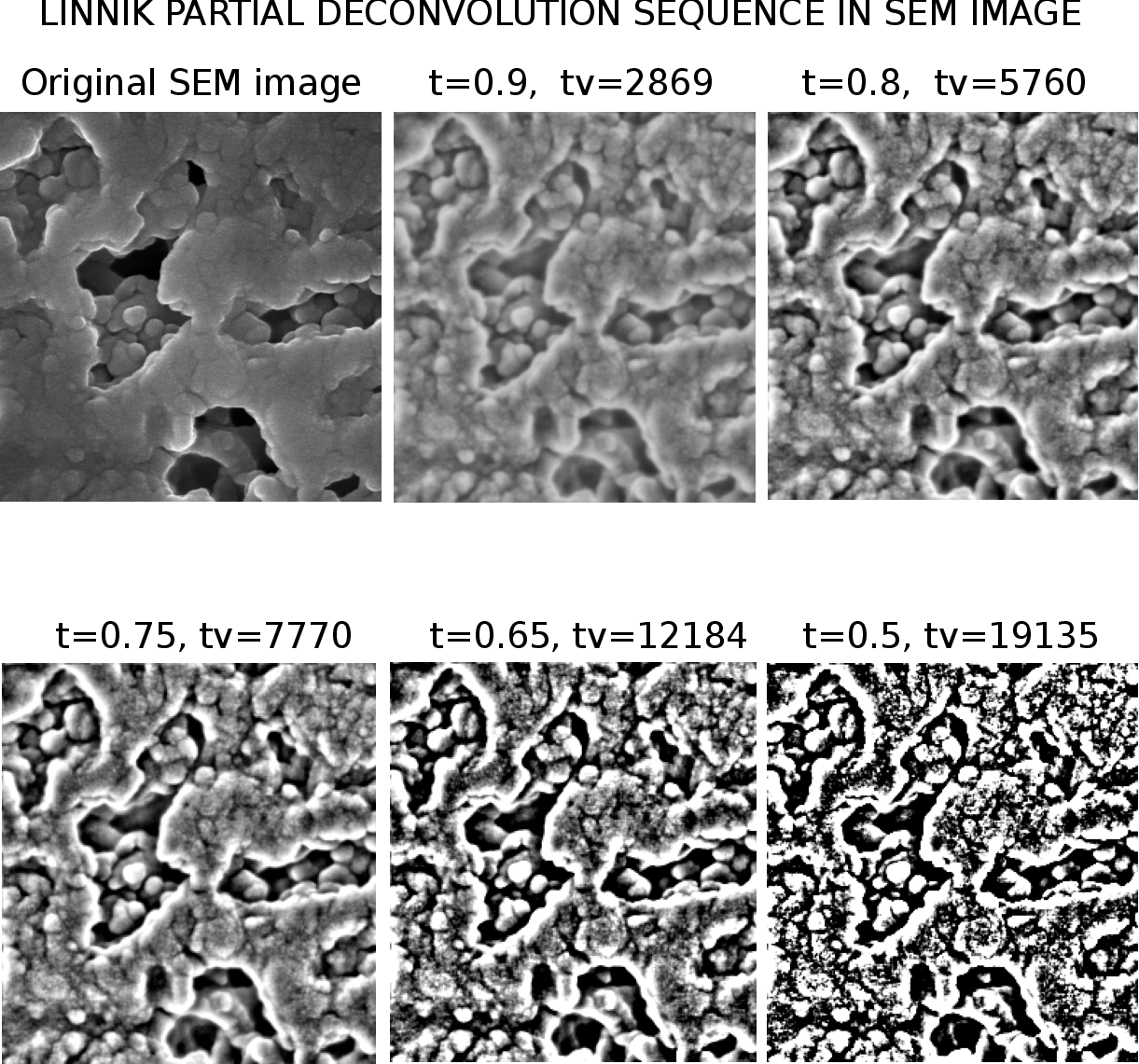}}
        \caption{Partial deconvolution sequence with $s=0.001,~K=300$. Best image is found at some $~\bar{t}~$ lying between $t=0.75$ and $t=0.8$. At smaller $t$ values,
 $\parallel w(.,t_n) \parallel_{tv}$ increases rapidly as serious noise develops. Above images are $1 \mu m$ HFW secondary electron images.}
\end{figure}

\section{Deconvolution by marching diffusion equations backward in time}
As was the case in \cite[Section 3]{carSEM}, the Linnik blind deconvolution problem is solved by marching the logarithmic diffusion equation,
$w_t=-\lambda\{log(I+\gamma(-\Delta)\} w$, backward in time from $t=1$ to $t=0$, using the {\em preconditioned} microscopy image $g(x,y)$ as data at $t=1$.  The {\em slow evolution} (SECB) constraint,
previously developed in \cite{carSEB}, is applied to stabilize the ill-posed backward computation. A complete discussion given in \cite[Section 3]{carAPEX} leads
to the partially deblurred Linnik SECB Fourier image $\hat{w}(\xi,\eta,t)$ defined as follows,
\begin{equation}
        \hat{w}(\xi,\eta,t)=\frac {\hat{h}(\xi,\eta,t) \hat{h}(\xi,\eta) \hat{g}(\xi,\eta)}{\hat{h}^2(\xi,\eta)+K^{-2}(1-\hat{h}(\xi,\eta,s))^2},
        \qquad 0 \leq t < 1,
        \label{eq:2.01}
\end{equation}
with suitably chosen positive constants $s,~ K$. Typical values for these constants might be $s=0.001,~ K=100$. An inverse Fourier transform in Eq.~(\ref{eq:2.01})
leads to $w(x,y,t)$, the partial deconvolution at time $t$. 
The above blind deconvolution procedure requires familiarity with microscopy images, as well as interactive search for useful values of
$s,~ K$. It may be helpful to produce a sequence of partially deconvolved images $w(x,y,t_n)$, for preselected decreasing values of $t_n$, with $1 > t_n > 0$.
The image $L^1$ norm, $\parallel w(.,t_n) \parallel_1$, as well as the
image total variation norm $\parallel w(.,t_n) \parallel_{tv}$,  where
\begin{equation}
	\parallel w(.,t_n) \parallel_1 = \int_{R^2} |(w(x,y,t_n)| dx dy, ~~~\parallel w(.,t_n) \parallel_{tv}=\int_{R^2} |\nabla w(x,y,t_n)| dx dy,
	\label{eq:2.02}
\end{equation}
may also be computed at each $t_n$. With a good choice of $s,~K$, it is typically found that $\parallel w(.,t_n) \parallel_1$ remains constant as $t_n \downarrow 0$, while 
$\parallel w(.,t_n) \parallel_{tv}$ increases systematically. However, rapidly increasing values of $\parallel w(.,t_n) \parallel_1$ often indicate development of noise.
A useful deblurred image might be found at $t_n=
\bar{t} \geq 0.7$, while smaller values for $t_n$ may produce images of lesser quality. This process is illustrated in Figure 2.3.
There are several distinct triples $(s,K,\bar{t})$ that can produce distinct useful deblurred images $w(x,y,\bar{t})$.
Examples of successful deblurred images, together with the parameters used in Eqs.~(\ref{eq:1.01}) and (\ref{eq:2.01}) for each image,
are given Figures 4.1 through 4.8.

\section{Concluding remarks}
Extensive discussions and comparisons are given in \cite{carIM}, between L\'{e}vy and Linnik point spread functions in
blind deconvolution of astronomical images.
While preconditioning was not needed in the astronomical images considered in \cite{carIM}, such preconditioning
plays a major role in electron microscopy. As illustrated in Figure 2.1, it is important to avoid oversmoothing the ADHE image.

Future applications of the above Linnik blind deconvolution procedure are contemplated for SEM imaging in biomedical, pharmaceutical,
and semiconductor contexts. Familiarity with each of these contexts will likely be necessary to obtain useful results.



\begin{figure}
	\centerline{\includegraphics[width=7.35in]{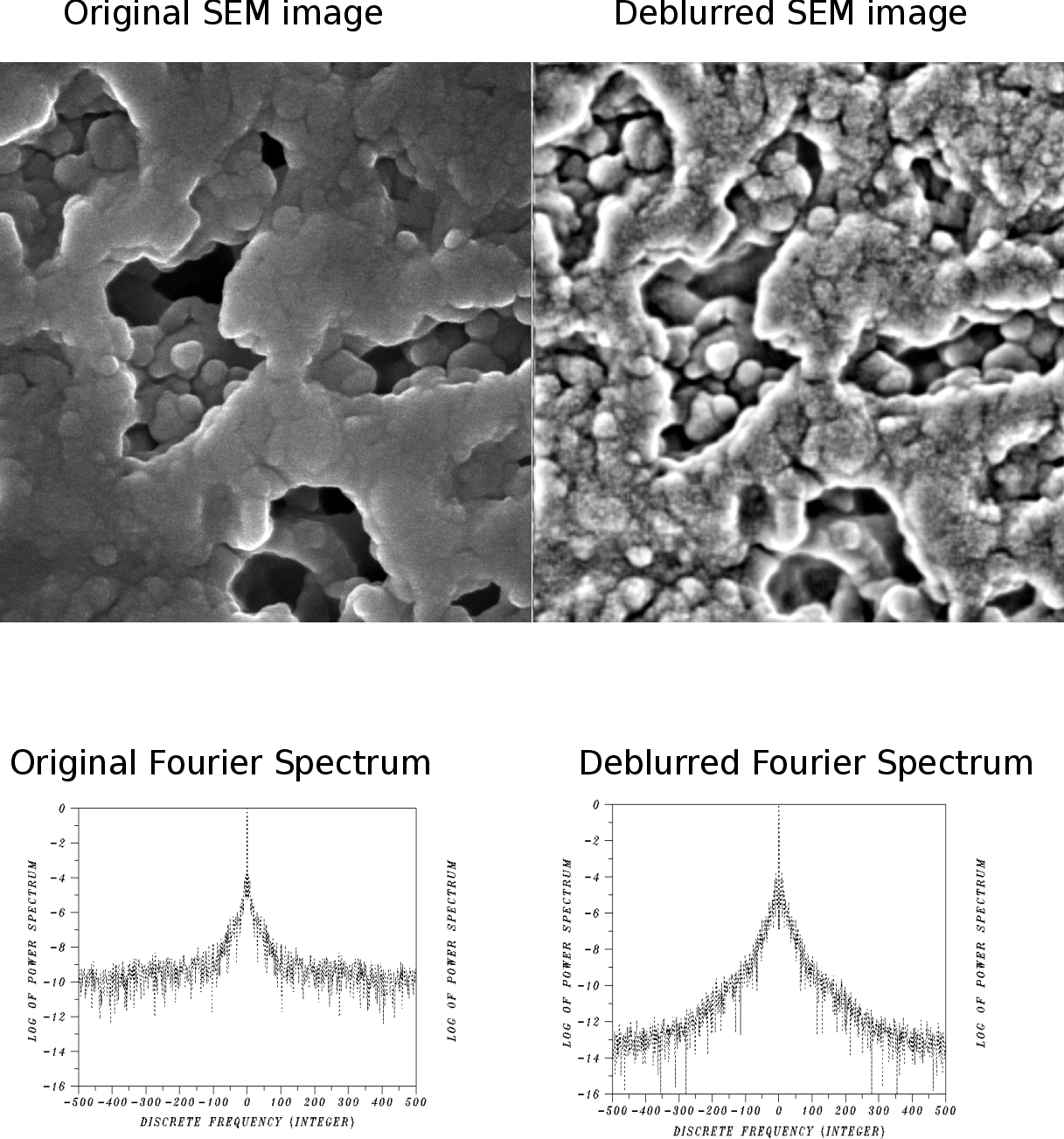}}
	\caption{$1 \mu m$ HFW images. Left: Secondary electron image of an etched glass sample showing weak surface details. Right: rich details after Linnik processing.
	\\
	Parameters in Eqs.~(\ref{eq:1.01}) and (\ref{eq:2.01}): $\lambda=0.969$,~$\gamma=1.64$,~$s=0.001,~K=300$,~$\bar{t}=0.77$.}
\end{figure}
\begin{figure}
	\centerline{\includegraphics[width=7.0in]{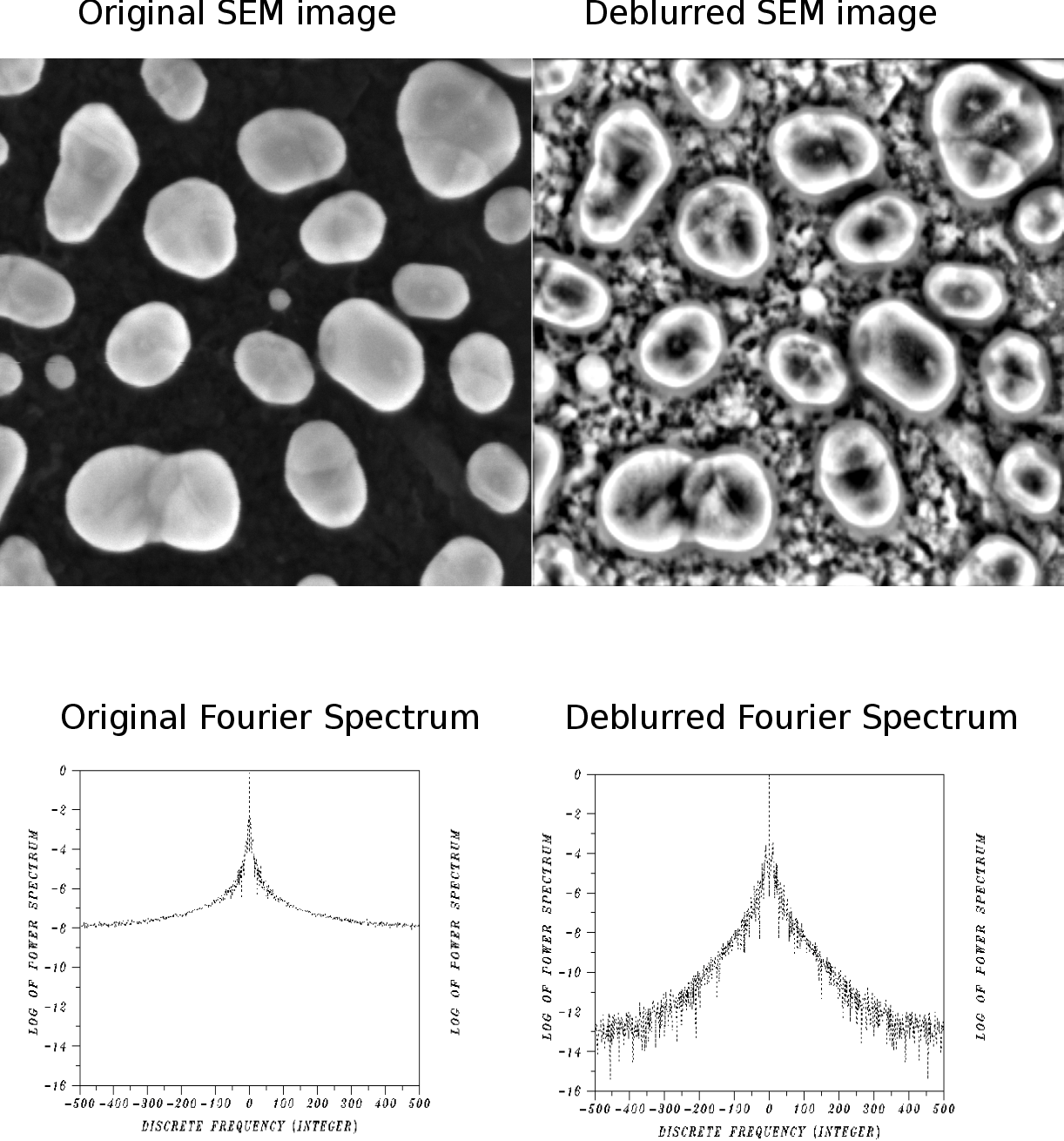}}
        \caption{$0.25 \mu m$ HFW images. Left: $0.5 nm$ resolution secondary electron image of Au nanoparticles on Si substrate with faint surface details. Right:
	rich details revealed after Linnik processing.
	\\
        Parameters in Eqs.~(\ref{eq:1.01}) and (\ref{eq:2.01}): $\lambda=0.989$,~$\gamma=1.226$,~$s=0.001,~K=500$,~$\bar{t}=0.77$.}
\end{figure}

\begin{figure}
	\centerline{\includegraphics[width=7.35in]{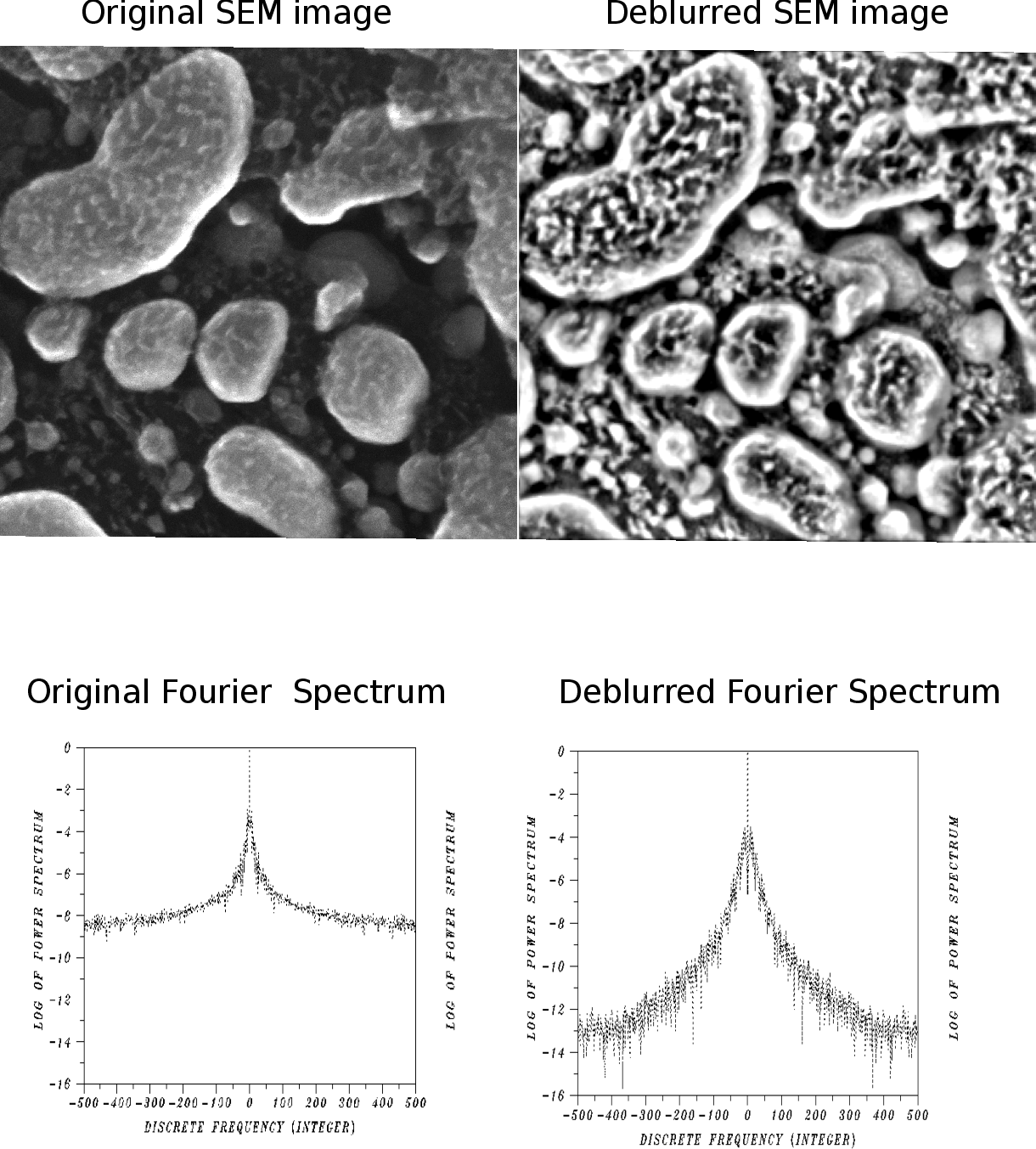}}
        \caption{$0.5 \mu m$ HFW images. Left: $0.7 nm$ resolution secondary electron image of Au nanoparticles on C substrate, showing surface details only on the gold particles.
	Right: rich details revealed on the C areas after Linnik processing.
	\\
        Parameters in Eqs.~(\ref{eq:1.01}) and (\ref{eq:2.01}): $\lambda=1.0,~\gamma=0.959$,~$s=0.001,~K=500$,~$\bar{t}=0.77$.}
\end{figure}

\begin{figure}
	\centerline{\includegraphics[width=7.35in]{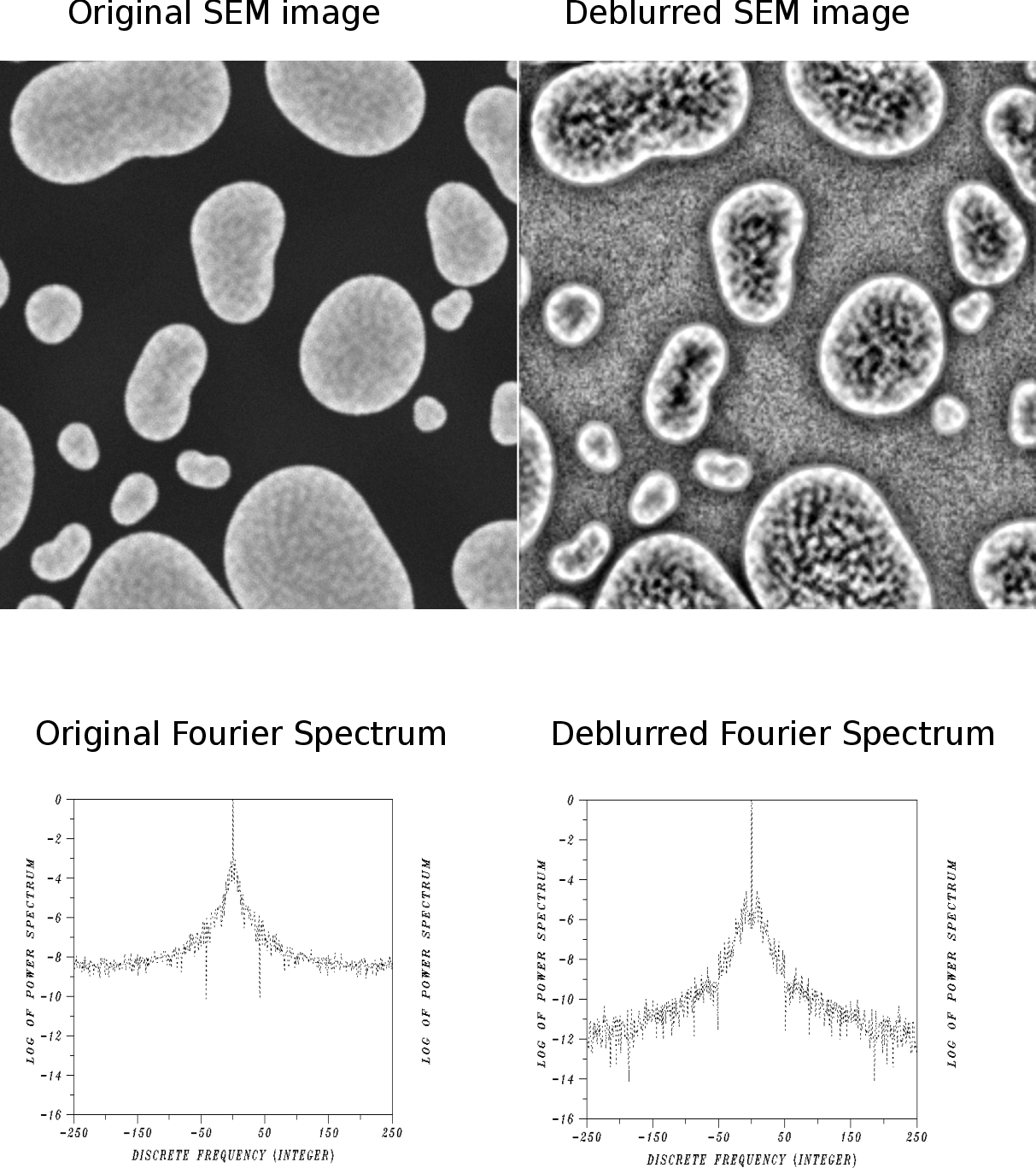}}
        \caption{$0.50 \mu m$ HFW images. Left: simulated secondary electron image of Au nanoparticles on Si substrate, showing surface details only on the gold particles.
	Right: gray level variations revealed on the Si areas after Linnik processing.
	\\
        Parameters in Eqs.~(\ref{eq:1.01}) and (\ref{eq:2.01}): $\lambda=0.993$,~$\gamma=1.06$,~$s=0.001,~K=500$,~$\bar{t}=0.8$.}
\end{figure}

\begin{figure}
	\centerline{\includegraphics[width=6.75in]{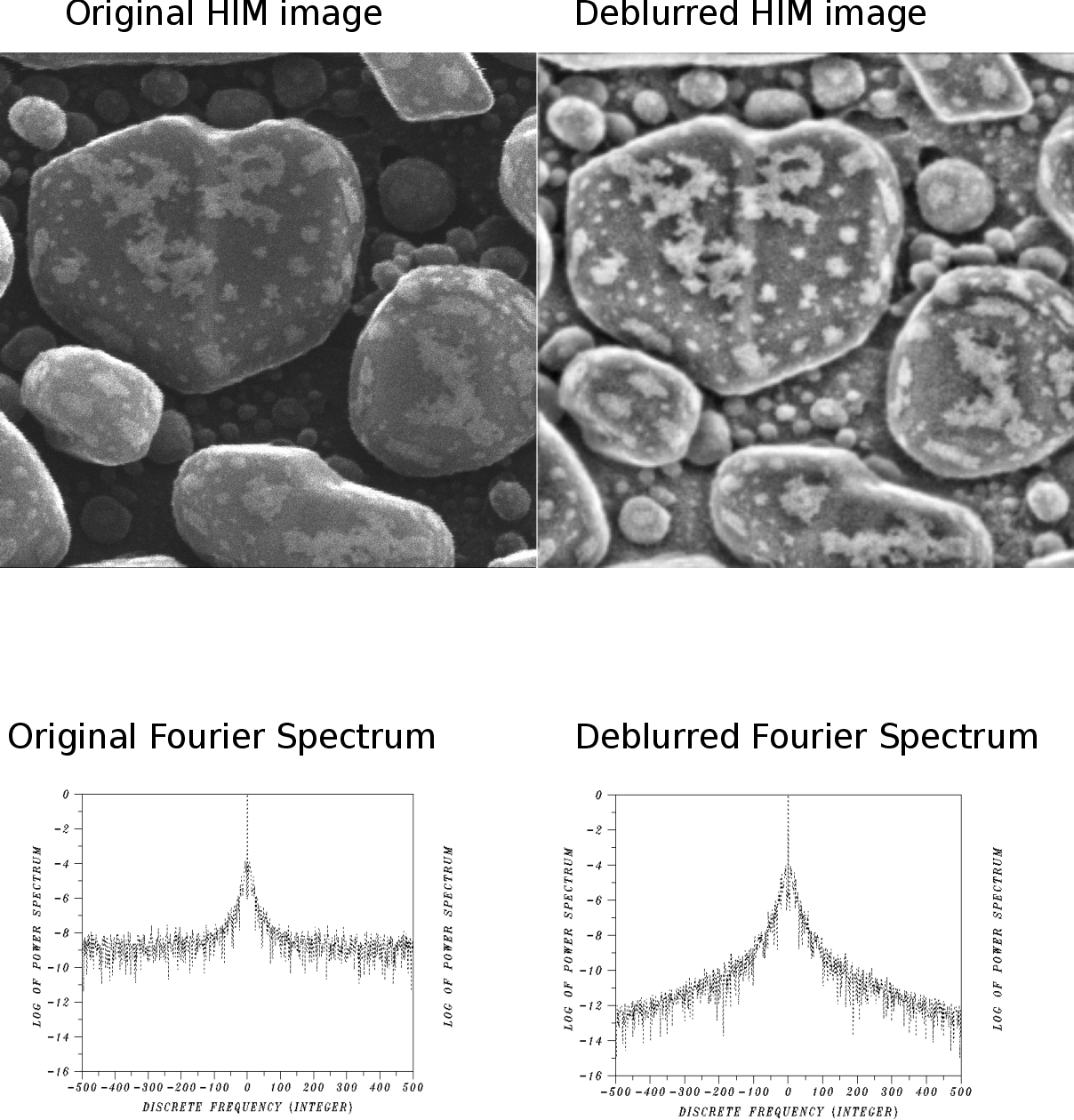}}
	\caption{$0.50 \mu m$ HFW HIM images. Left: 0.7nm resolution secondary electron image
	of Pt-decorated Au nanoparticles on C substrate, showing surface details on gold particles only. Right: rich details revealed on the C areas after Linnik processing.
	\\
        Parameters in Eqs.~(\ref{eq:1.01}) and (\ref{eq:2.01}): $\lambda=1.0$,~$\gamma=0.367$,~$s=0.001,~K=500$,~$\bar{t}=0.77$.}
\end{figure}

\begin{figure}
	\centerline{\includegraphics[width=7.25in]{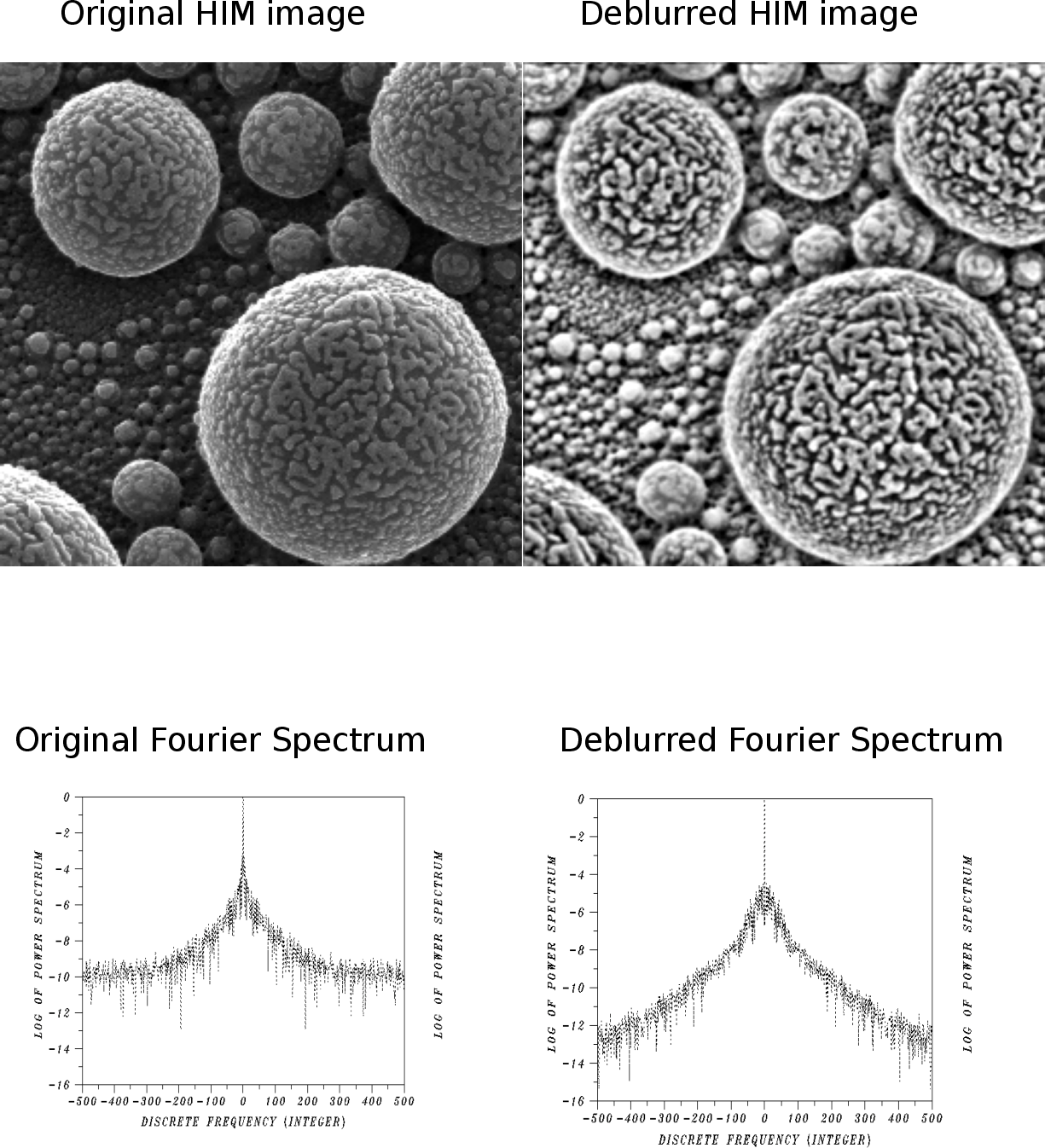}}
	\caption{$5.0 \mu m$ HFW HIM images. Left: high resolution Pt-decorated tinballs on Si substrate, showing surface details on the gold particles. 
	Right: rich details revealed in the dark areas after Linnik processing.
	\\
        Parameters in Eqs.~(\ref{eq:1.01}) and (\ref{eq:2.01}): $\lambda=1.0$,~$\gamma=0.367$,~$s=0.001,~K=500$,~$\bar{t}=0.73$.}
\end{figure}

\begin{figure}
	\centerline{\includegraphics[width=7.35in]{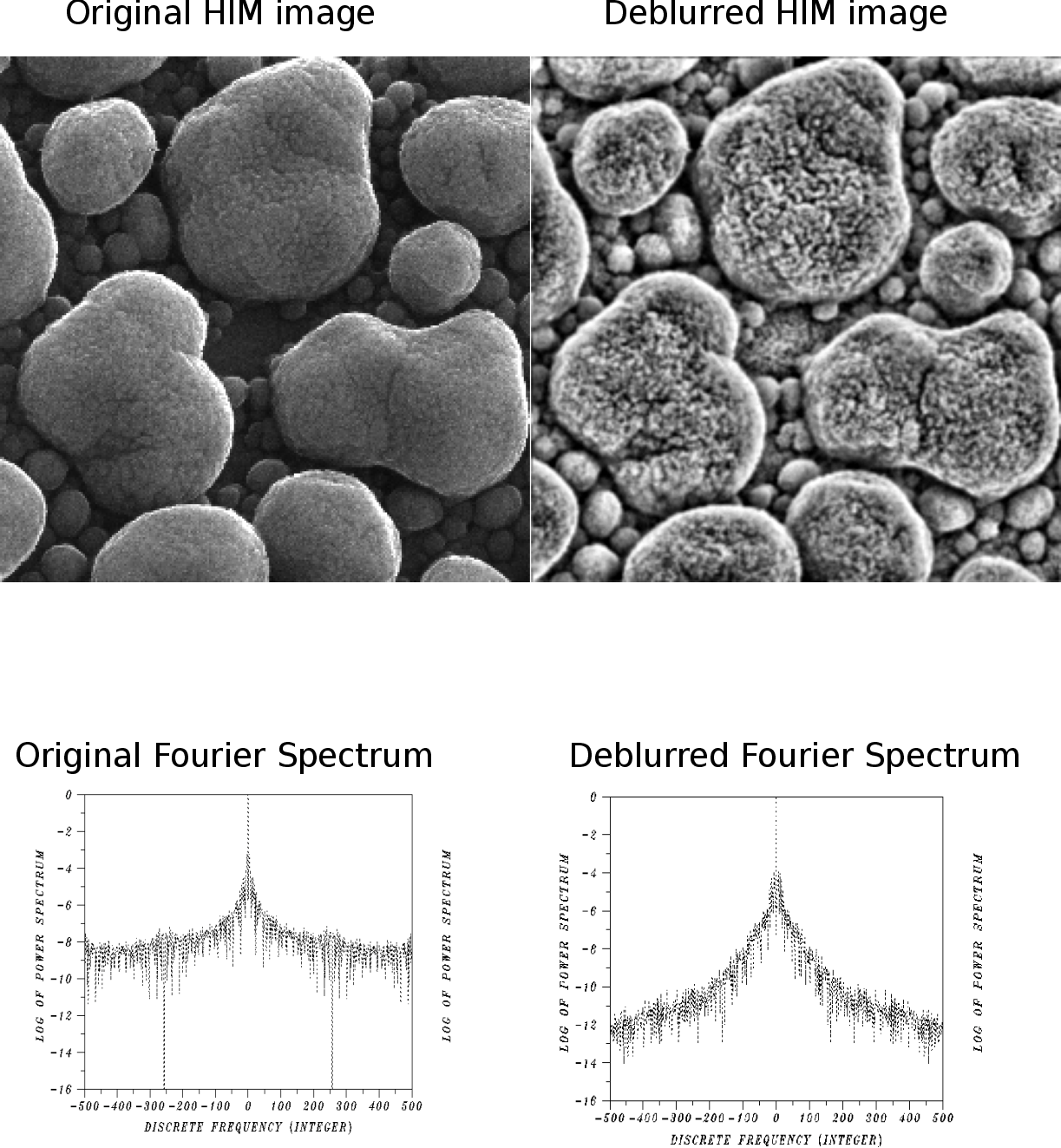}}
	\caption{$2.0 \mu m$ HFW HIM images. Left: 0.7nm resolution secondary electron image of Au nanoparticles on C substrate, showing surface details on the gold particles only. Right: rich details revealed after Linnik processing.
	\\
        Parameters in Eqs.~(\ref{eq:1.01}) and (\ref{eq:2.01}): $\lambda=1.0$,~$\gamma=0.367$,~$s=0.001,~K=500$,~$\bar{t}=0.77$.}
\end{figure}

\begin{figure}
	\centerline{\includegraphics[width=7.35in]{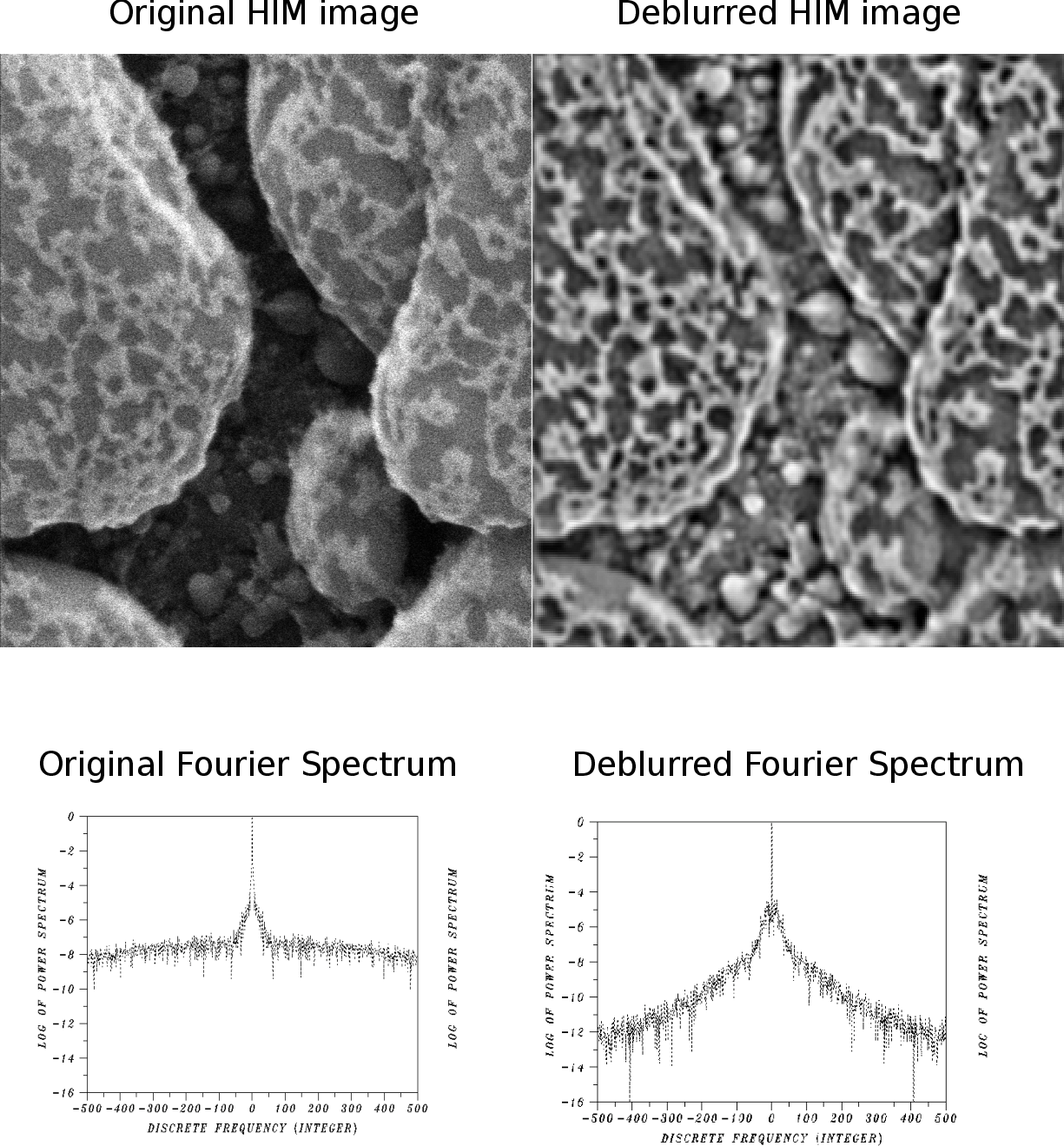}}
	\caption{$0.5 \mu m$ HFW HIM images. Left: 0.7nm resolution secondary electron image of Pt-decorated Au nanoparticles on C substrate, showing surface
	details on the gold particles only. Right: rich details revealed after Linnik processing.
	\\
        Parameters in Eqs.~(\ref{eq:1.01}) and (\ref{eq:2.01}): $\lambda=1.0$,~$\gamma=0.367$,~$s=0.001,~K=500$,~$\bar{t}=0.77$.}
\end{figure}

\ \\

\end{document}